\documentclass[aps,prl,twocolumn]{revtex4}
\usepackage{graphicx}
\RequirePackage{times}
\usepackage[dvips,unicode,colorlinks,linkcolor=blue,citecolor=blue,urlcolor=blue]{hyperref}
\begin{document}
\author{A. V. Savin$^{1}$, E. A. Zubova$^{1}$, L. I. Manevitch$^{1}$}

\title{Can low-frequency breathers exist in a quasi-1D crystal?}
\affiliation{ $^1$ Institute of
Chemical Physics RAS, 4 Kosygin str., Moscow 119991, Russia }

\begin{abstract}
We investigate a quasi-1D crystal: 2D system of coupled linear
chains of particles
with strong intra-chain and weak inter-chain interactions.
Nonlinear dynamics of one of these chains when the rest of them
are fixed is reduced to the well known Frenkel-Kontorova (FK)
model. Its continuum limit, sine-Gordon (sG) equation, predicts
two types of soliton solutions: topological solitons and breathers.
It is known that the quasi-1D topological solitons exist also in a
2D system of coupled chains and even in a 3D model of a polymer
crystal. Numerical simulation shows that the breathers
inherent to the FK model do not exist in the system of chains. The
effect changes scenario of kink-antikink collision at small
velocities: it always results only in intensive phonon radiation
while kink-antikink recombination in the FK model results in
long-living low-frequency sG breather creation. The reason of the
difference is that the model of coupled chains catches 'acoustic'
part of phonon spectrum of a crystal while the FK model does not.
Low-frequency SG breathers in a crystal intensively emit resonant
'acoustic' phonons and come to ruin.
\end{abstract}
\maketitle

The Frenkel-Kontorova (FK) model (a linear chain of harmonically
coupled particles on the sine substrate) \cite{FK} is the the most
commonly used and comprehensively investigated (see monographs
\cite{Kiv1,Kiv2}) 1D model of a crystal. In the case of weak
substrate potential (weakly discrete system) it seems to be
especially appropriate for polymer crystals:
quasi-1D topological soliton-like excitations predicted by its
continuum limit, sine-Gordon (sG) equation, exist in
a 2D system of coupled chains \cite{s1,s2} and even in a 3D
model of a polymer crystal (see, for example, \cite{Zubova-tension}).
SG equation is the only nonlinear wave equation of
type
%%%%%%%%%%%%%%%%%%%%%%%%%%%%%%%%%%%%%%%%%%%%%%%%%
\begin{equation}
u_{tt}-u_{xx}+g(u)=0 \label{f1}
\end{equation}
%%%%%%%%%%%%%%%%%%%%%%%%%%%%%%%%%%%%%%%%%%%%%%%%%
which possesses also one-parametric family of exact solutions in the
form of low-frequency breathers \cite{sine-unique}. Frequencies of the
breathers fill the gap between zero and the minimal frequency in
phonon spectrum $\Omega_{FK}(0)$. If the breather frequency
approaches $\Omega_{FK}(0)$, the breather amplitude tends to zero,
and the breather width -- to infinity ('phonon' limit). If the
breather frequency tends to zero, the breather approaches a full
kink-antikink profile.

Numerical simulations \cite{sd-sGb, wd-sGb}
show that in the FK model the sG breathers survive, and, although
lose the energy due to resonances of odd multiples to the breather
frequency with phonon frequencies, have their lifetime long enough
even in the case of strong discreteness (more than hundred periods
when the third harmonic to its frequency becomes higher than the
upper phonon band edge \cite{sd-sGb}). In the case of weak
discreteness losses of energy are hardly perceptible \cite{wd-sGb}.

In connection with studying real physical quasi-1D systems such as
long Josephson junctions and quasi-1D ferromagnets there emerged
many works treating behavior of sG breathers under action of
perturbations breaking exact integrability: dissipative and
diverse conservative terms (see \cite{Scott-79, Remoissenet-86},
and \cite{Kivshar-Malomed-87} and references therein). Analytical
treatment of the problem is possible as perturbation in the inverse
scattering transform \cite{Scott-79} or as multiple-scale
asymptotic expansion \cite{Remoissenet-86} in the limit of high
breather frequencies, and one can obtain some estimates in general
case \cite{Kivshar-Malomed-87}. As one can easily predict, the
breather lifetime proved to be long if perturbation is small. In
nonintegrable models with substrate potentials sufficiently
different from the sine function breather-like long-living
nonlinear excitations are observed numerically ($\phi^4$ -
\cite{Kudryavtsev}, double sG, square well potential -
\cite{Ablowitz}). For the $\phi^4$ model it is shown
\cite{Segur-phi4} that the radiation rate of a small-amplitude
'breather' lies beyond all orders in asymptotic expansion.

All this allows one to look on such breathers as being
'elementary excitations' in a crystal, together with kinks and
antikins (topological solitons) and phonons. This implies that the
breathers can noticeably contribute to thermodynamic properties of
a crystal \cite{Bishop-b} and even must be used in
phenomenological approaches to sG thermodynamics instead of
phonons \cite{Sasaki, Chung}. We show that this conclusion based
on analysis of the FK model (a chain on a substrate) is not
valid for a more realistic model of a crystal: a system of coupled
chains.

\begin{figure}[t]
\begin{center}
\includegraphics[angle=0, width=1.0\linewidth]{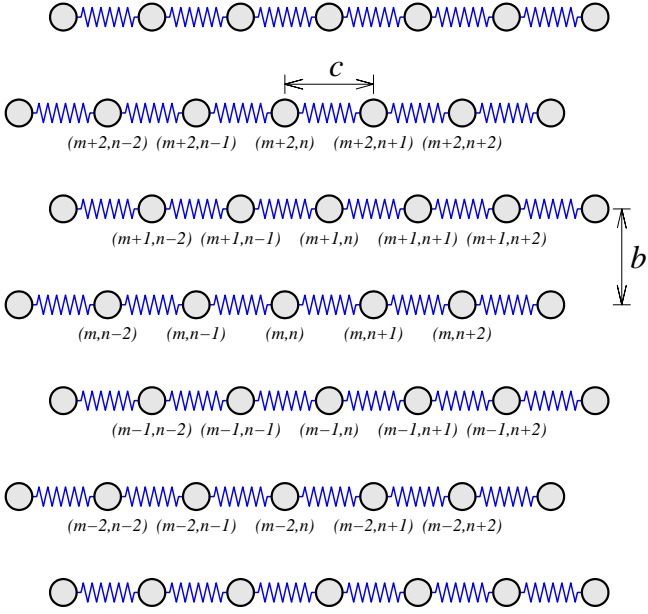}
\end{center}
\caption{\label{fig1}\protect\small 2D model of quasi-1D system: intra-chain
interactions are much stronger than inter-chain ones (weak discreteness limit).
We choose unit length so that $c=1$ in the equilibrium ground state of the
crystal.}
\end{figure}

Let us take a system of coupled linear chains of (classical)
particles (fig.\ref{fig1}).
 %---------------------------- Fig. 1 ------------------------------------
%---------------------------- Fig. 1 ------------------------------------
To catch the main physical meaning of the model it is enough to
allow inter-chain interactions only between particles of
the nearest neighboring chains. Then Hamiltonian of the system
is written as
\begin{eqnarray}
H=\sum_{m,n}\{\frac12\dot{u}_{m,n}^2
+\frac12(u_{m,n+1}-u_{m,n}+c-c_0)^2
\nonumber \\
+\sum_{j=-\infty}^{+\infty} U(r_{m,n;j})\},
\label{full-Hamiltonian}
\end{eqnarray}
where the dot denotes time derivative, $c_0$ is the period of a
separate chain, $c$ - the longitudinal period of the crystal,
$u_{m,n}$ is longitudinal deviation of the particle $(m,n)$ from
its equilibrium position (shown in fig.\ref{fig1}) in the
crystal (we keep transversal deviation $y_{m,n}=0$), the potential
$U(r_{m,n;j})$ describes interaction of the $n$-th particle in the
$m$-th chain with the $(n+j)$-th particle in the $(m+1)$-th chain,
$r_{m,n;j}$ being the distance between the particles
$$
r_{m,n;j}=\{[(j-(-1)^m/2)c+u_{m+1,n+j}-u_{m,n}]^2+b^2\}^{1/2}.
$$
The ground state of the system ($u_{m,n}=\dot{u}_{m,n}=0$)
has the energy per particle
$$
E(b,c,c_0)=(c-c_0)^2 + \sum_{j=-\infty}^{+\infty}
U(R_{j}),
$$
where $R_{j}=[b^2 + c^2(j+1/2)^2]^{1/2}$ is the distance between the $n$-th and
the $(n+j)$-th particles of the $m$-th ($m$ is odd) and
$(m+1)$-th chains. Equilibrium
values of $b$ and $c$ minimize the expression at given $c_0$. Equivalently, if
one chooses $c$ as unit length, one can find equilibrium values of $c_0$ and
$b$. Hereafter we imply that the crystal is initially in its equilibrium ground
state with $c=1$.

The present 2D model of quasi-1D crystal was first introduced in \cite{s1}. One
can take into account also transversal displacements of particles \cite{s2}.
The model allows existence and propagation of quasi-1D topological soliton-like
excitations. In \cite{s1,s2} have been used the Morse potential of particle
interactions $U(r_{m,n;j})$ as very suitable for numerical calculations. Here
we exploit the more physical Lennard-Jones potential (truncated):
\begin{equation}
U(r)=\varepsilon\left(\frac{r_0}{r}\right)^6\left[\left(\frac{r_0}{r}
\right)^6-2\right]f(r)
\end{equation}
where the truncation function $f(r)=\{1-\tanh[\mu(r-d_0)]\}/2$
($\mu\sim 1$, $d_0\gg r_0$) is introduced for convenience of
numerical calculations. It allows one to avoid taking into account
interactions of the particles placed one from another farther than
$r \approx d_0$.
%---------------------------- Fig. 2------------------------------------
\begin{figure}[t]
\begin{center}
\includegraphics[angle=0, width=1.0\linewidth]{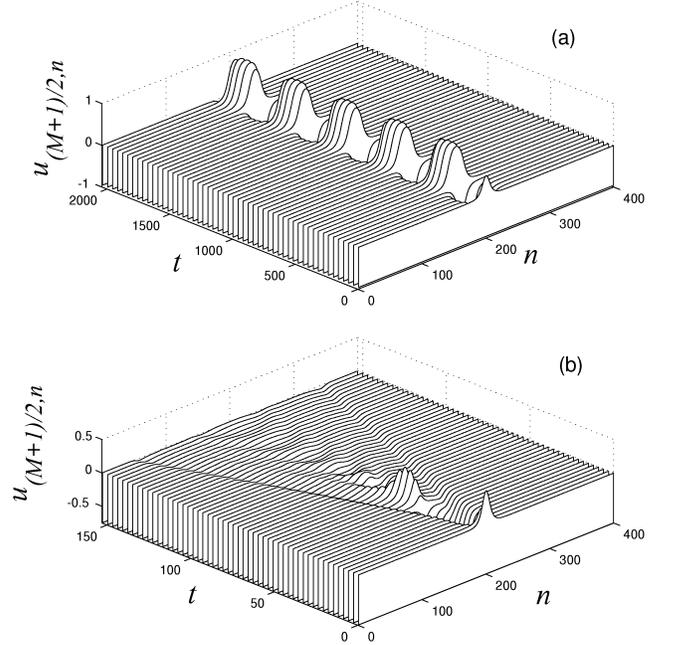}
\end{center}
\caption{\label{fig2}\protect\small Low-frequency sG breather in the crystal
($\epsilon=0.007$, $M=51$, $N=400$): oscillations in the FK model (the case of
immobile neighboring chains) (a) and degradation in the full model of
interacting mobile chains (b). We show the displacements $u_{26,n}$ of the 26th
chain containing the breather in successive time moments $t$. Breather
frequency is $\omega_b=0.0157$.} \end{figure}
%---------------------------- Fig. 2 ------------------------------------

It is known \cite{Zubova_01} that if the equilibrium distance
between particles $r_0$ falls into the interval $0.91<r_0<\infty$
shape of the substrate potential is close to the sine function. We
have chosen $r_0=1.67$ ($d_0=20$, $\mu=2$) because it corresponds
to model of polyethylene crystal with 'united atoms'
\cite{Zubova-tension, Balabaev_95}. At this value of $r_0$ the
substrate generated by immobile neighbors
\begin{equation}
V(u)=2\sum_{j=-\infty}^{+\infty}[U(r_j(u))-U(R_j)], \label{potV}
\end{equation}
where $r_j(u)=[b^2+(u+j+1/2)^2]^{1/2}$ (note that $R_j=r_j(0)$),
is the sine function accurate within $0.1\%$: $V(u)\simeq
\epsilon[1-cos(2\pi u)]$, where substrate amplitude
$\epsilon=0.1757\varepsilon$. So dynamics of one chain when the
rest of them are fixed is reduced to the well known FK model:
\begin{equation}
H_{FK}=\sum_{n}\{\frac12\dot{u}_n^2+\frac12(u_{n+1}-u_n)^2+V(u_n)\}.
\label{FK}
\end{equation}
The width of a static kink in the model of polyethylene crystal
with 'united atoms' (about 30 periods) coincides with the width of
a static kink in our model of coupled chains if the intensity of
inter-chain interactions $\varepsilon=0.0007$. We have also
considered the cases of stronger interactions
$\varepsilon=0.007,~0.07$. The first two cases correspond to limit
of weak discereteness. In the last case the sound velocity in
transversal direction is equal to one in longitudinal direction
(see table \ref{tab1}). When the chains are assembled into the
crystal the transversal equilibrium period appears to be
$b=1.5666$ independent on $\varepsilon$. Only $c_0$ is
$\varepsilon$-dependent.
\begin{table}[b]
  \centering
  \caption{Dependence of the longitudinal $s_x$ and transversal $s_y$
sound velocities, characteristic frequencies $\omega_{max}$ and
$\Omega_{FK}(0)$ on the value of the parameter $\varepsilon$.}\label{tab1}
\begin{tabular}{c|cccc}
\hline
 $\varepsilon$ & $s_x$ & $s_y/b$ & $\omega_{max}$ & $\Omega_{FK}(0)$ \\
\hline
0.07  & 0.7530 & 0.4916 & 2.1174 & 0.6952\\
0.007 & 0.9781 & 0.1555 & 2.0120 & 0.2197\\
0.0007& 0.9978 & 0.0492 & 2.0012 & 0.0694\\
\hline
\end{tabular}
\end{table}

The system of equations of motion for the quasi-1D crystal
takes the form:
\begin{eqnarray}
\ddot{u}_{m,n}=-\frac{\partial H}{\partial u_{m,n}},
\label{equations} \\
n=0,\pm1,\pm2,...,~~m=0,\pm1,\pm2,... \nonumber
\end{eqnarray}
with the Hamiltonian (\ref{full-Hamiltonian}).
In numerical simulations we considered the dynamics of
a bounded rectangular fragment of the crystal
$(1\le n\le N,~1\le m\le M)$
with fixed boundary conditions in both directions.

We have compared behavior of a sG breather in the FK model
(all the chains are kept immobile except one - with number
$m=(M+1)/2$ - containing the breather) and in the model of coupled
chains (all the chains are mobile).

\begin{figure}[t]
\begin{center}
\includegraphics[angle=0, width=1.0\linewidth]{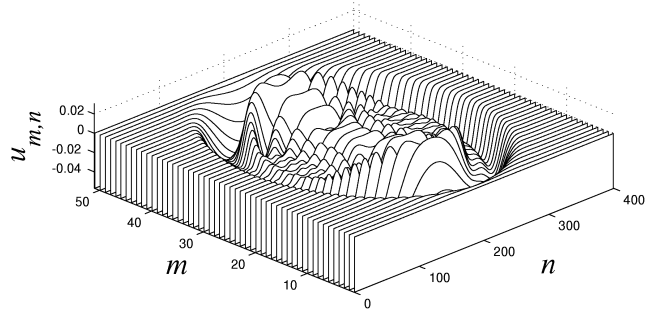}
\end{center}
\caption{\label{fig3}\protect\small
State of the crystal ($M=51$, $N=400$) of coupled chains
($\epsilon=0.007$) resulting from destruction of low-frequency
($\omega_b=0.0157$) sG breather placed onto the 26th chain. The
displacements $u_{m,n}$ are shown in the time moment $t=150$.}
\end{figure}

Numerical simulation shows that the breather in the FK model
enjoys regular stable oscillations for a very long time
(fig.\ref{fig2} (a)). The situation changes drastically if we
allow all the chains in the crystal to move. The breather quickly
comes to ruin. Its lifetime is less than two its periods
(fig.\ref{fig2} (b)). The effect is observed by all three values
of the inter-chain interactions $\varepsilon=0.07,0.007,0.0007$.
The destruction results from the intensive emission of phonons
into the neighboring chains (see fig.\ref{fig3}). Energy of
the breather spreads to all the particles. So one can conclude
that the low-frequency sG breathers are absent in the model of
coupled chains.
%---------------------------- Fig. 3------------------------------------
%---------------------------- Fig. 3 ------------------------------------

\begin{figure}[t]
\begin{center}
\includegraphics[angle=0, width=1.0\linewidth]{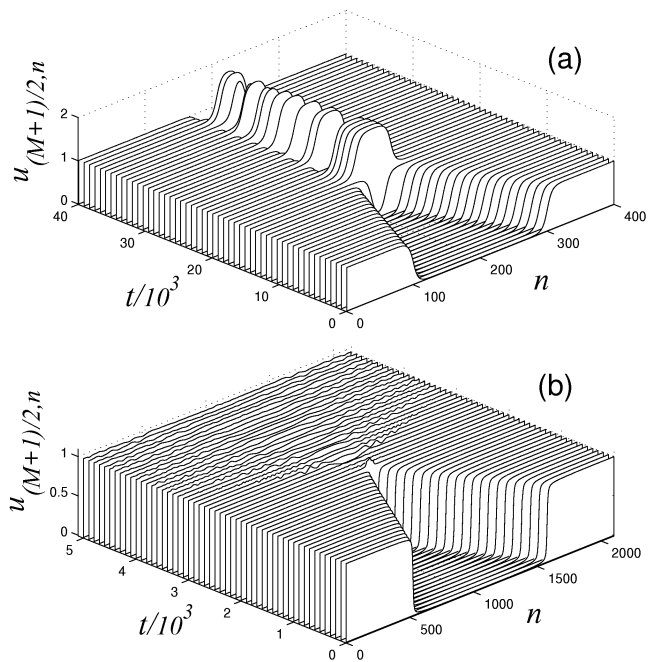}
\end{center}
\caption{\label{fig4}\protect\small
Kink-antikink recombination in a chain of the crystal
($\varepsilon=0.007$) having
immobile (a) and mobile (b) neighboring chains. Collision of
solitons in the FK model (soliton velocity
$s=0.005$) results in creation of a low-frequency sG breather.
Collision of solitons in the model of all moving chains
(soliton velocity $s=0.25$) results in
intensive radiation of phonons.}
\end{figure}

With this in mind, one can suppose the difference between the two
models under study in scenario of kink-antikink recombination
when they collide with small velocities. Indeed,
in the FK model (with the sine potential as well as with the double
sine or square well ones) kink and antikink can form
oscillating breather-like state, their energy remaining for a long
time localized \cite{Ablowitz} -- see fig. \ref{fig4} (a),
while in the model of coupled chains their energy scatters at ones
with phonons -- see fig.\ref{fig4} (b).
%---------------------------- Fig. 4 ------------------------------------
%---------------------------- Fig. 4 ------------------------------------

\begin{figure}[t]
\begin{center}
\includegraphics[angle=0, width=1.0\linewidth]{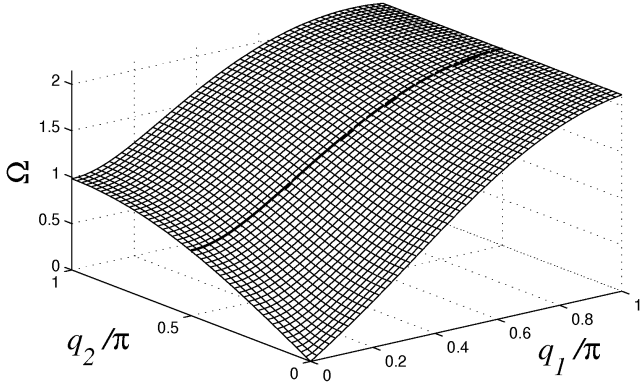}
\end{center}
\caption{\label{fig5}\protect\small Dispersion surface
$\omega=\Omega (q_1,q_2)$ for the model of coupled chains. The
model parameters $r_0=1.67$, $\varepsilon=0.07$ . The curve on the
surface is the dispersion curve for corresponding FK model
(approximation of immobile chains) $\omega=\Omega (q_1,\pi/2)$. }
\end{figure}

The reason of the effect observed seems to be that the two models of a crystal
possess qualitatively different phonon spectra. Namely, the FK model has a gap
between $\omega=0$ and the lower edge of the spectrum $\Omega_{FK}(0)$ while in
the model of coupled chains - like in a genuine crystal - the minimal possible
frequency is $\omega_{min}=0$. Let us show it. As we have chosen particles
numeration not coinciding with one based on translation of the crystal cell,
phonon modes have the more complicated form:
\begin{eqnarray}
u_{2m,n}&=&A\exp i[q_1n+q_22m-\omega t], \nonumber \\
u_{2m+1,n}&=&A\exp i[q_1(n-1/2)+q_2(2m+1)-\omega t], \label{anzatz}\\
&& n=0,\pm1,\pm2,...,~~~m=0,\pm 1,\pm 2,...~. \nonumber
\end{eqnarray}
where $A\ll 1$, and $q_1,~q_2\in [0,\pi]$. Substituting the anzatz
(\ref{anzatz}) into the linearized system of equations
(\ref{equations}) with imposed periodic boundary conditions
in both directions one can obtain the dispersion equation
\begin{eqnarray}
\Omega(q_1,q_2)=\{2(1-\cos q_1)\nonumber \\
+ 4\sum_{j=0}^{+\infty} K_{j}[1-\cos((j+1/2)q_1)\cos q_2]\}^{1/2},
\label{dispersion}
\end{eqnarray}
where rigidities are $ K_{j}=U''(R_{j})(j+1/2)^2/R_{j}^2+U'(R_{j})b^2/R_{j}^3.
$ Values of the rigidities are in direct proportion to the parameter of
inter-chain interation $\varepsilon$; for $\varepsilon=0.07$ they are
$K_{0}=0.199$, $K_{1}=-0.061$, $K_{2}=-0.014$, $K_{3}= -0.002$. We have
presented the plot of the dispersion surface for $\varepsilon=0.07$ in
fig.\ref{fig5} together with the dispersion curve for the corresponding FK
model.
%---------------------------- Fig. 5------------------------------------
%---------------------------- Fig. 5 ------------------------------------

Dispersion equation (\ref{dispersion}) gives the minimal
$\omega_{min}=\Omega(0,0)=0$ and the maximal
$$
\omega_{max}\equiv\Omega(\pi,q_2)=2 (1+\sum_{j=0}^{+\infty}K_j)^{1/2}
$$
possible frequencies,
the velocity of longitudinal long phonons
$$
s_x=\lim_{q_1\rightarrow 0}\Omega (q_1,0)/q_1=
[1+\sum_{j=-\infty}^{+\infty}(j+1/2)^2K_{j}]^{1/2},
$$
and the velocity of transversal long phonons
$$
s_y= b \lim_{q_2\rightarrow 0}\Omega (0,q_2)/q_2= b
(\sum_{j=-\infty}^{+\infty}K_{j})^{1/2}.
$$
The cut of the dispersion surface (\ref{dispersion})
at $q_2=\pi/2$ produces the dispersion curve for the corresponding
FK model:
$$
\Omega_{FK}(q)=\Omega(q,\pi/2)= \{2[1-\cos(q)] +
2\sum_{j=-\infty}^{+\infty}K_j\}^{1/2}.
$$
Now the minimal frequency is $\Omega_{FK}(0)=(2\sum_j K_j)^{1/2}>0$. Dependence
of the velocities $s_x$, $s_y$, and the frequencies $\omega_{max}$ and
$\Omega_{FK}(0)$ on the parameter of inter-chain interactions $\varepsilon$ is
presented in the Table \ref{tab1}.

So, with such a phonon spectrum, the model of interacting chains
can not possess even approximate low-frequency breather solutions.
Their frequencies fall into phonon band, the resonance interaction
between a breather and a phonon takes place and the breather
energy is transmitted to the phonons, resulting in quick breather
degradation. And evidently it is the case in real (for example,
polymer) crystals.

The authors thank the Russian Foundation of Basic Research (awards
04-02-17306 and 04-03-32119) for financial support. One of the
authors (E.A.Z.) acknowledges also substantial help of the Russian
Science Support Foundation.


\begin{thebibliography}{99}

\bibitem{FK} T.A. Kontorova, Ya.I. Frenkel, Zh. Eksp. Teor.
Fiz. {\bf 8}, 89 (1938) (in Russian); T.A. Kontorova, Ya.I. Frenkel, Zh. Eksp.
Teor. Fiz. {\bf 8}, 1340 (1938) (in Russian).
\bibitem{Kiv1} O.M. Braun, Yu.S. Kivshar, Physics Reports {\bf 306}, 1
(1998).
\bibitem{Kiv2} O.M. Braun, Yu.S. Kivshar, The Frenkel-Kontorova Model:
Concepts, Methods, and Applications, Springer-Verlag  (2004).
\bibitem{s1}
P.L. Christiansen, A.V. Savin, A.V. Zolotaryuk, Phys. Rev. B, {\bf 57}, 13564
(1998).
\bibitem{s2}
A.V. Savin, J.M. Khalack, P.L. Christiansen, A.V. Zolotaryuk, Phys. Rev. B,
{\bf 65}, 054106 (2002).
\bibitem{Zubova-tension}
E.A. Zubova, N.K. Balabaev, L.I. Manevitch, Zhur. Eksper. Teor. Fiz. (Moscow)
{\bf 115}, 1063 (1999) [J. Exp. Theor. Phys., {\bf 88}, 586 (1999)].



\bibitem{sine-unique}
S. Kitchenassamy, Communications on Pure and Applied Mathematics {\bf 44}, 789
(1991).
\bibitem{sd-sGb}
R. Boesch, M. Peyrard, Physical Review B {\bf 43}, 8491 (1991).
\bibitem{wd-sGb}
S.V. Dmitriev, T. Shigenari, K. Abe, A.A. Vasiliev, A.E. Miroshnichenko,
Computational Materials Science {\bf 18}, 303 (2000).
\bibitem{Scott-79}
A.C.Scott, Phys. Scripta {\bf 20}, 509 (1979).
\bibitem{Remoissenet-86}
M.Remoissenet, Physical Review B {\bf 33}, 2386 (1986).
\bibitem{Kivshar-Malomed-87}
Y.S. Kivshar, B.A. Malomed, Europhysics Letters {\bf 4}, 1215 (1987).
\bibitem{Kudryavtsev}
A.E. Kudryavtsev, JETP Letters {\bf 22}, 82 (1975).
\bibitem{Ablowitz}
M.J. Ablowitz, M.D. Kruskal, J.F. Ladik, SIAM Journal of Applied Mathematics
{\bf 36}, 428 (1979).
\bibitem{Segur-phi4}
H. Segur, M.D. Kruskal, Physical Review Letters, {\bf 58}, 747 (1987).
\bibitem{Bishop-b}
A.R. Bishop, Journal of Physics A: Math. and Gen. {\bf 14}, 1417 (1981).
\bibitem{Sasaki}
K.Sasaki, Physical Review B {\bf 33}, 2214 (1986).
\bibitem{Chung}
S.G. Chung, International Journal of Modern Physics B, {\bf 8}, 2447 (1994).
\bibitem{Zubova_01}
E.A. Zubova, Zhur. Eksper. Teor. Fiz. (Moscow) {\bf 120}, 1027 (2001)
[J. Exp. Theor. Phys., {\bf 93}, 895 (2001)].
\bibitem{Balabaev_95}
N.K. Balabaev, O.V. Gendelman, M.A. Mazo, L.I. Manevitch,
Zh. Phys. Chim. {\bf 69}, 24 (1995) (in Russian).
\end{thebibliography}
\end{document}